\documentclass[a4paper,12pt]{article}
\usepackage{amsmath,amssymb,amsfonts,color,graphicx,cite,color,feynarts}
\usepackage{caption}
\usepackage[labelfont=bf,textfont={sl,bf}]{subfig}
\newcommand{\A}{{\cal A}}

\newcommand{\Rel}{\textrm{Re\,}}
\newcommand{\Img}{\textrm{Im\,}}

\newcommand{\be}{\begin{equation}}
\newcommand{\ee}{\end{equation}}
\newcommand{\bea}{\begin{eqnarray}}
\newcommand{\eea}{\end{eqnarray}}
\newcommand{\ba}{\begin{array}}
\newcommand{\ea}{\end{array}}
\newcommand{\beal}{\begin{aligned}}
\newcommand{\eeal}{\end{aligned}}
\newcommand{\nn}{\nonumber}

\newcommand{\fr}{\frac}
\newcommand{\de}{\delta}

\newcommand{\al}{\alpha}
\newcommand{\la}{\lambda}

\newcommand{\bet}{\beta}

\newcommand{\pa}{\partial}
\newcommand{\bc}{\begin{center}} 
\newcommand{\ec}{\end{center}}
\newcommand{\Ga}{\Gamma}
\newcommand{\ep}{\epsilon}

\newcommand{\Si}{\Sigma}

\newcommand{\ti}{\tilde}

\def\slashepi{\epsilon_i\kern -.720em {/}}
\def\slashpi{p_i\kern -.600em {/}}

\def\slashc{c\kern -.400em {/}}
\def\slashp{p\kern -.400em {/}}
\def\slashq{q\kern -.450em {/}}
\def\slashL{L\kern -.450em {/}}
\def\slashcl{\cl\kern -.600em {/}}
\def\slashr{r\kern -.450em {/}}
\def\slashk{k\kern -.500em {/}}
\def\slashep{\epsilon\kern -.450em {/}}
\def\slashpbar{\bar{p}\kern -.450em {/}}
\def\slashD{D\kern -.650em {/}}

\newcommand{\ie}{{\it i.e.\;}}

\newcommand{\CP}{\text{CP }}

\newcommand{\TeV}{\text{TeV}}
\newcommand{\GeV}{\text{GeV}}

\newcommand{\DRb}{\overline{\text{DR}}}
\newcommand{\bfZ}{\textbf{Z}}
\newcommand{\opi}{\text{1PI}}

\newcommand{\tree}{\text{tree}}

\evensidemargin \oddsidemargin
\allowdisplaybreaks

\begin{document}
\begin{titlepage}
\vspace*{0.1cm}\rightline{MPP-2010-118}

\vspace{3cm}
\begin{center}

{\large \textbf{CP violating asymmetry in $H^\pm\to W^\pm h_1$ decays }}\\

\vspace{.5cm}

W. Hollik and D.T. Nhung

\vspace{4mm}

{\it Max-Planck-Institut f\"ur Physik (Werner-Heisenberg-Institut), \\
D-80805 M\"unchen, Germany}

\end{center}
\vspace*{0.1cm}
\begin{abstract}
The CP violating asymmetry from the decay rates 
 $H^\pm\to W^\pm h_1$ of charged Higgs bosons into the lightest neutral
Higgs boson and a $W^\pm$ boson is calculated and discussed in 
the complex MSSM. 
The contributions from all complex phases are considered, 
especially from the 
top-squark trilinear coupling, which induces a large contribution to the 
CP asymmetry.
\end{abstract}
\end{titlepage}

\section{Introduction}

In the Minimal Supersymmetric Standard model with complex parameters 
(complex MSSM), new sources of CP violation are associated with the phases
of the soft-breaking parameters
and of the Higgsino-mass parameter $\mu$.
Through loop contributions, CP violation also enters the 
Higgs sector, which is CP conserving at lowest order
(see for example~\cite{Accomando:2006ga} for a detailed study and references).
As a consequence, the $h, H$ and $A$ neutral Higgs bosons in general
mix and form the 
neutral mass eigenstates $h_{1,2,3}$ with both CP even and odd properties,
giving rise to
CP-violation in suitable observables,  like decay rates
of charged particles.
An interesting decay mode is given by the charged Higgs boson decays
$H^-\to W^- h_1$ and $H^+\to W^+ h_1$,
where the asymmetry between the decay rates
is a CP-violating observable. 
A first calculation was done in~\cite{Christova:2003hg}, 
studying the CP asymmetry as derived from the phases of the trilinear
$\tilde{\tau}$ coupling, $A_{\tau}$, and of $M_1$, yielding asymmetries
of the order  $10^{-2}$; contributions from the quark/squark sector
were not included.

In this paper we extend the calculation of~\cite{Christova:2003hg} 
including contributions
from all physical phases in the general complex MSSM
with minimal flavor violation, in particular 
from $A_t$ and $A_b$, which enter through Feynman diagrams with stops
and sbottoms involving large Yukawa couplings,
further enhanced by the color factor. 
We show the results from the complete set of one-loop diagrams, 
including besides 
the Higgs self energies all the loop contributions to the  
$H^\pm\to W^\pm h_1$ vertex, which at lowest order is in general 
suppressed by a factor $\cos(\beta-\alpha)$. 

The paper is organized as follows. 
We first outline in section~\ref{sec:convention}, 
the structure of the  complex MSSM neutral Higgs bosons.
In section 3, we indicate the calculation of the \CP decay rate
asymmetry. 
A discussion of  the results follows in section 4,
and conclusions in section~5.

\section{The Higgs sector of the complex MSSM}
\label{sec:convention}

In the MSSM, 
\CP violation arises from the Yukawa sector and the soft SUSY-breaking 
sector through complex couplings. 
Physical phases are the phase of the trilinear couplings $A_f$,  
of the higgsino parameter $\mu$,  of the gaugino mass parameters
$M_i$ ($i$ = 1,2,3), 
\bea 
 A_f =|A_f|e^{i\phi_f}\,,\quad \mu=|\mu|e^{i\phi_\mu}\,, 
\quad M_{i}=|M_{i}|e^{\phi_{i}}, 
\eea
and the CKM phase as in the Standard Model.
The phase of the CKM matrix has a very small impact on the \CP asymmetry 
considered here and is neglected in the following.
At tree level, the complex  SUSY phases 
enter the mass matrices of squarks, sleptons, charginos and 
neutralinos.
In the Higgs sector, \CP violation effects enter only at the loop level. 
\\ \\
\underline{\textbf{Tree level:}}\\
Using the conventions of~\cite{Frank:2006yh}, we write the 
two Higgs doublets in the form 
\bea 
\label{doublets}
&&{\cal H}_1 = \left( \ba{c} v_1 + \fr 1{\sqrt 2}(\phi_1 - i\chi_1) \\
                          -\phi_1^-\ea\right)  \,, 
\quad {\cal H}_2 = \left( \ba{c} \phi_2^+\\
                          v_2+ \fr 1{\sqrt 2}(\phi_2 + i\chi_2)\ea\right),
\eea
with the vacuum expectation values $v_1, v_2$, 
yielding the ratio $\tan\beta = v_2/v_1$. 
The  mass eigenstates are related to the field components
in ~(\ref{doublets})
by  unitary matrices, for the neutral Higgs case given by
\bea \left(\ba{c} h\\H\\A\\G\ea\right) = 
\left(\ba{cccc}-\sin\alpha & \cos\alpha&0&0\\ \cos\alpha & 
\sin\alpha&0&0\\ 0&0&-\sin\beta_n&
\cos\bet_n\\ 0&0& \cos\bet_n&\sin\bet_n\ea\right)
\left(\ba{c} \phi_1\\\phi_2\\\chi_1\\\chi_2\ea\right), \eea
and for the charged Higgs fields by
\bea \left(\ba{c} H^\pm\\G^\pm\ea\right) =
 \left(\ba{cc} -\sin\bet_c&\cos\bet_c
\\ \cos\bet_c&-\sin\bet_c\ea\right) \left(\ba{c} \phi_1^\pm\\
\phi_2^\pm\ea\right),
 \eea 
with  $\bet_n=\bet_c=\bet$.
At the tree level, 
the Higgs potential conserves CP, hence the CP-even states 
$h, \, H$ do not mix with the CP-odd states $A$. 
\\ \\
\underline{\textbf{Higher order:}}\\
Through the nonvanishing \CP phases in the loop contributions
mixing between $h,H$ and $A$ occurs. Moreover, there is mixing
of the neutral Higgs bosons with $G$ and $Z$, 
but they yield only sub-leading two-loop 
contributions to the Higgs boson masses, see e.g.~\cite{Frank:2006yh}. 
The lowest-order mass eigenvalues
 $m_h$, $m_H$ and $m_A$ are different from the pole masses.
The loop-corrected masses (pole masses)
of the neutral Higgs are obtained via the poles of the propagator matrix,
\bea 
\Delta_{hHA} 
             = - \big[ \hat{\Ga}_{hHA}(p^2) \big] ^{-1},
\label{eq:delta}
\eea
with 
\bea
\beal \hat{\Ga}_{hHA}(p^2) &= i\big[p^2 -  M(p^2)\big]\,,\\
 M(p^2)& =
  \left(\ba{ccc}
    m_h^2 - \hat{\Si}_{hh}(p^2) & - \hat{\Si}_{hH}(p^2) & - 
\hat{\Si}_{hA}(p^2) \\
    - \hat{\Si}_{hH}(p^2) & m_H^2 - \hat{\Si}_{HH}(p^2) & - 
\hat{\Si}_{HA}(p^2) \\
    - \hat{\Si}_{hA}(p^2) & - \hat{\Si}_{HA}(p^2) & m_A^2 - 
\hat{\Si}_{AA}(p^2)
  \ea\right),
\eeal\eea
where $\hat\Si_{ij}$ ($i,j =h, H, A$) are the renormalized self-energies
in the scheme of~\cite{Frank:2006yh}, which treats 
the renormalization of the Higgs fields and of $\tan\beta$ 
according to the $\DRb$ prescription. 
In general, the three poles are complex and  written as 
\bea 
{\cal M}_{h_a}^2 = M_{h_a}^2 - i M_{h_a} \Ga_{h_a} \,,\quad a=1,2,3,
\eea
where $M_{h_a}$ are the loop-corrected masses with the convention
\bea M_{h_1} < M_{h_2} < M_{h_3},\eea
and $\Ga_{h_a}$ are the corresponding total decay widths. 
The mass of the charged Higgs-boson is chosen
as an input parameter and is renormalized on-shell. 
Again, there is also mixing between $H^\pm$
and $G^\pm, W^\pm$ at one-loop order, which has to be taken 
 into account in processes with external charged Higgs bosons.

\section{Decay widths and \CP asymmetry}

The \CP violating asymmetry in the charged-Higgs decay into a $W$-boson and the 
lightest neutral Higgs, $h_1$, is defined in the following way
\bea 
\de_{\CP} = \fr {\Ga(H^-\to W^-h_1) - \Ga(H^+\to W^+h_1)}
 {\Ga(H^-\to W^-h_1) + \Ga(H^+\to W^+h_1)} 
\label{eq:asymmetry}
\eea 
in terms of the individual partial decay widths $\Ga(H^\pm\to W^\pm h_1)$.
Writing the decay amplitudes as follows, 
\bea
&& {\cal A}(H^\pm\to W^\pm h_1) =  
 \big( \ep_\la \cdot p_{H^\pm} \big) \;
 {\cal M}(H^\pm\to W^\pm h_1) 
\eea
with the $W$ polarization vectors $\ep_\la$ and the $H^\pm$ momentum
$p_{H^\pm}$, the decay widths integrated over the 2-particle phase space 
and summed over the $W$ helicities $\la$
are obtained in the form
\bea 
\Ga(H^\pm\to W^\pm h_1) = R_2 \cdot | M_{H^\pm\to W^\pm h_1}|^2\,,
 \label{eq:decaywidth}
\eea
with
\bea 
R_2 &=& 
\fr {\la^{3/2}(M^2_{H^\pm}, M^2_W, M^2_{h_1})}{64\pi M_{H^\pm}^3M_W^2}, \quad
\la(x,y,z) = x^2+y^2+z^2-2xy-2xz-2yz \, .
\eea

\begin{figure}
 \begin{center}
\input{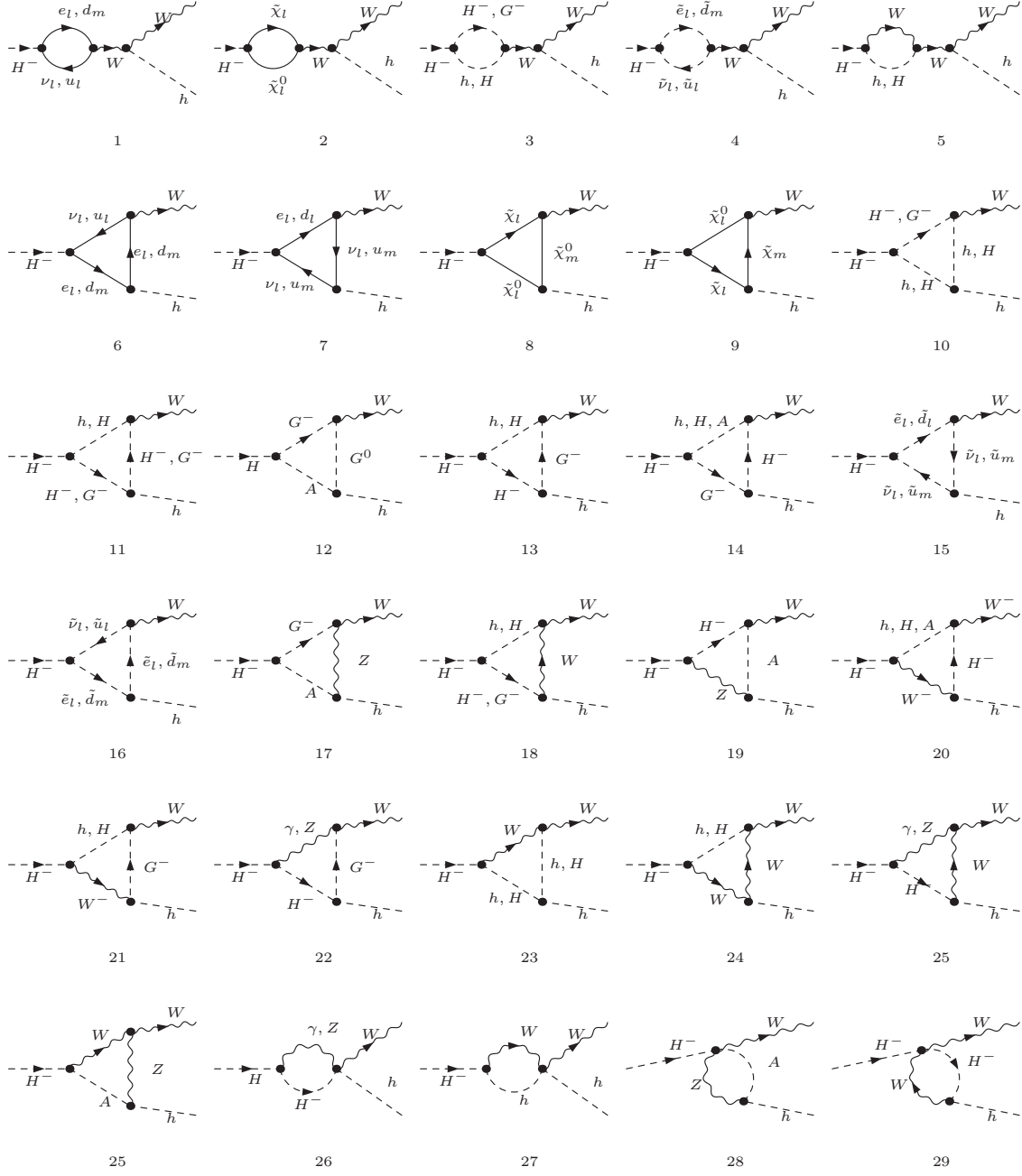}
\caption{One-loop Feynman diagrams contribute to $\de M_h$ }
\label{fig:triangle}
\end{center}
\end{figure}

\noindent
The decay amplitude at higher order  
can be written in the following way,
\bea 
\beal {\cal M}_{H^\pm\to W^\pm h_1} &= \sqrt{Z_{H^-H^+}}\big[\bfZ_{11} 
\big(M^{\tree}_{H^\pm\to W^\pm h}+ \de M_{H^\pm\to W^\pm h}\big)\\
&+ \bfZ_{12} M^{\tree}_{H^\pm\to W^\pm H} + 
\bfZ_{13} M^{\tree}_{H^\pm\to W^\pm A}\big],
\eeal
\label{eq:vertex}
\eea
with the tree-level expressions $M^{\tree}$ given by
(with $s_W = \sin\theta_W$)
\bea 
M^{\tree}_{H^\pm\to W^\pm h} =\fr{e\cos(\bet -\al)}{s_W}, 
\quad M^{\tree}_{H^\pm\to W^\pm H} =-\fr{e\sin(\bet -\al)}{s_W},
 \quad M^{\tree}_{H^\pm\to W^\pm A} =\pm i\fr{e}{s_W},
\eea
the charged-Higgs wave function renormalization $\sqrt{Z_{H^-H^+}}$,
the neutral-Higgs wave function renormalization factors 
${\bf Z}_{kl}$, and
\bea 
\de M_h \equiv \de M_{H^\pm\to W^\pm h}=\de M^{\opi}_{H^\pm\to W^\pm h} 
+ \de M^{G,W mix}_{H^\pm\to W^\pm h}
\eea
which summarize the residual 1PI-irreducible contributions to the 
3-point vertex function and the mixing of  $H^\pm$ with $G^\pm$ and $W^\pm$.  
The  Feynman diagrams contributing to this 
term at the one-loop level are shown in figure \ref{fig:triangle}. 
There is no explicit wave function renormalization for the $W$ boson,
since the $W$ propagator has been renormalized on-shell
yielding residue~$= 1$.

The Higgs fields $H^\pm$, $h_1$ and $\tan\bet$ are renormalized 
in $\DRb$ scheme. 
The correct on-shell
properties of the S-matrix element involving external neutral Higgs 
bosons are
ensured by 
the inclusion of the wave function renormalization factors 
summarized in the matrix ${\bf Z}$, as given in~\cite{Frank:2006yh}: 
\bea 
\bfZ = \begin{pmatrix}
             \sqrt{Z_h} & \sqrt{Z_h}Z_{hH}&\sqrt{Z_h}Z_{hA}\\
              \sqrt{Z_H}Z_{Hh}& \sqrt{Z_H}& \sqrt{Z_H}Z_{HA}\\
               \sqrt{Z_A}Z_{Ah}& \sqrt{Z_A}Z_{AH}&\sqrt{Z_A}
            \end{pmatrix},
\eea
where ($i,j = h, H, A$)
\begin{align}
 Z_i& = \fr 1 {\left(\fr{i}{\Delta_{ii}(p^2)}\right)^\prime(M_i^2)},\nn\\
Z_{ij} &= \fr {\Delta_{ij}(p^2)}{\Delta_{ii}(p^2)}\bigg|_{p^2=M_i^2},
\end{align}
involving the elements $\Delta_{ij}$ of the 
the propagator matrix $\Delta_{hHA}$ in~(\ref{eq:delta}).

For the charged Higgs boson, 
the wave function  renormalization is derived from 
\bea 
Z_{H^-H^+} = \big[1 + \Rel \fr{\pa}{\pa p^2}
\hat{\Si}_{H^-H^+}\big]^{-1}_{\big|_{p^2=M^2_{H^\pm}}},
\eea
with the $\DRb$-renormalized self-energy
$\hat{\Si}_{H^-H^+}$.
At one-loop order we get 
\bea 
Z_{H^-H^+} &\simeq& 1 - \Rel \fr{\pa}{\pa p^2}\hat{\Si}_{H^-H^+}
\big|_{p^2=M^2_{H^\pm}} 
 \, \equiv\,  1-\delta Z_{H^-H^+} \, , \nonumber \\
\sqrt{Z_{H^-H^+}} &\simeq& 1 - \frac{1}{2}\, \delta Z_{H^-H^+}.
\label{eq:RWfatorH}
\eea
The factor $Z_{H^-H^+}$ is IR-divergent.
We regularise the IR-divergence in the one-loop expanded version with the 
help of a small photon mass, to be canceled by including 
real photon bremsstrahlung.

\setlength{\unitlength}{0.030mm}

Substituting the amplitude~(\ref{eq:vertex}) into the 
expression~(\ref{eq:decaywidth}), 
one obtains the decay width, denoted 
as $\Ga_{\bfZ}^{(0+1+2)}$ later in the paper. 
Keeping the ${\bf Z}$ factors in the squared amplitude 
is justified since they contain also the leading higher-order terms
which correspond to the effective-potential approximation.
In the squared one-loop amplitude, we also keep 
the term involving $\de M_h^2$.
This term can play an important role at large value of $M_{H^-}$,\ie
$M_{H^-}\ge M_{\ti t_1}+ M_{\ti b_1}$, where the decay channel into $\ti t_1$
 and $\ti b_1$ is open, while it is negligible at lower $M_{H^-}$. 
The inclusion of this term while neglecting other two-loop contributions is 
consistent in perturbation theory, since the tree-level vertex function
 $M^{\tree}_h\sim\cos(\bet -\al)\sim M_{Z}^2/M_{H^-}^2$ 
 goes to near zero at large $M_{H^-}$.
The IR divergence at the one-loop level is canceled 
by adding the real photon radiation contribution. 
An IR-divergence in the squared one-loop term is avoided 
by taking only the  (s)top/(s)bottom diagrams which 
are IR finite and give the dominant 
contributions, as checked in~\cite{santi}.

\begin{figure}
 \begin{center}
\input{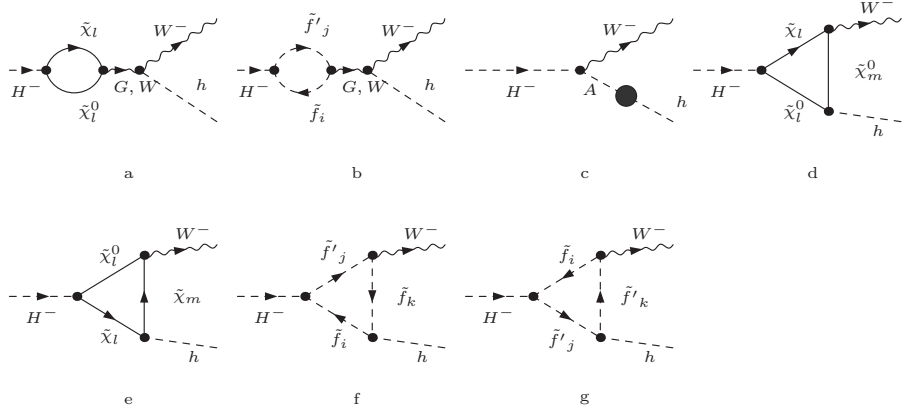}
\caption{Feynman diagrams contain weak phases which contribute to the \CP 
rate asymmetry}
\label{fig:asymmetry}
\end{center}
\end{figure}

In practice, there are two ways to compute the \CP asymmetry: (i)
to compute both decay widths of $H^-\to W^- h_1$ and of the CP-conjugate
process $H^+\to W^+ h_1$ and then using the definition~(\ref{eq:asymmetry});
(ii) to compute separately  the CP-violating  and
 the CP-invariant contributions to the decay 
$M_{H^-\to W^- h_1}$ and then taking their ratio.
The CP-violating term comes from the imaginary part
of the complex couplings (together with the imaginary part of the 
loop integrals), 
while the CP-invariant term is from the real part.
Therefore the CP-violating term change sign, but the CP-invariant term
does not when going from $H^-\to W^-h_1$ to $H^+\to W^+h_1$.
Hence, one can identify the Feynman diagrams shown in figure 
\ref{fig:asymmetry} as those contributing to the CP-violating part. 


We have performed our calculation in the two ways, with perfect
agreement. 
The full result for $\de_{CP}$ is obtained when both the 
numerator and denominator of the asymmetry~(\ref{eq:asymmetry}) 
are computed with the inclusion of higher order terms.
This is different with the approximation used
in Ref. \cite{Christova:2003hg} where the numerator is computed at 
strict one-loop order and the denominator is tree-level like, and is
necessary since 
in specific case the process is loop dominated,
as we will illustrate in the numerical analysis.

For comparison with other approximations, 
we introduce the  following notations for decay width:
\begin{itemize}
 \item 
The improved Born approximation 
for  the  decay width $\Ga_{\bfZ}^{(0)}$ with the $\bfZ$ 
factors taken into account:
\bea 
\Ga_{\bfZ}^{(0)}= R_2\cdot \big|\sum_{i}\bfZ_{1i}M_i^{\tree} \big|^2,
\quad 
i=h,H,A.
\label{eq:improveborn}
\eea 
\item The one-loop improved decay width $\Ga_{\bfZ}^{(0+1)}$ that  does not
include $\de M_h^2$:
\bea 
\Ga_{\bfZ}^{(0+1)}= R_2\cdot \bigg[\big|\sum_{i}\bfZ_{1i}M_i^{\tree}\big|^2
+ 2\sum_{i}\big| \bfZ_{11}^*\bfZ_{1i}M_i^{\tree}(\de M_h-\fr12M_h^{\tree}
\delta Z_{H^-H^+})^*\big|\bigg]. 
\label{eq:improveloop}
\eea
\end{itemize}

\setlength{\unitlength}{0.030mm}

\section{Numerical analysis}
\subsection{Calculational  frame work}
We have used  \texttt{FeynArts~3.4} \cite{Kublbeck:1990xc} to generate the
Feynman diagrams. In order to include the relevant counterterms, 
we have adapted the MSSM model file in FeynArts. The amplitudes are further
evaluated by \texttt{FormCalc~6.0} and the one-loop integrals are 
computed with the library  \texttt{LoopTools~2.4} \cite{Hahn:1998yk}. 
All the dependent couplings and masses of internal lines are computed with 
 tree-level relations. The mass of 
the external neutral Higgs is calculated by using 
\texttt{FeynHiggs~2.6.5} \cite{vanOldenborgh:1989wn}. In 
\texttt{FeynHiggs~2.6.5}, one has possibility to include various important 
two-loop contributions to the renormalized self-energies. We have included 
the full-phase-dependent $\al_s\al_t$ corrections and the ($\al_s\al_b,
\,\al_t\al_t,\, \al_t\al_b$) corrections which are interpolated in the complex
phases. Therefore, the most up-to-date higher-order renormalization factors
 $\bfZ_{ij}$ are used in our calculation.  

We should mention the problem of 
normal threshold singularities when  $M_{H^\pm}$ approaches
the production threshold of two scalar particles, for instance 
up and down squarks.  
Following~\cite{Kniehl:2000kk} and references therein, this problem
can be overcome by using complex masses for the relevant unstable particles. 
In our case, the kinematical threshold of top and bottom squarks
is concerned. This singularity appears 
in the renormalization factor of the charged Higgs boson, $\delta Z_{H^-H^+}$,
in particular in the derivative of two-point functions,
which we treat according to the substitutions
\bea 
\fr{d}{dp^2} B(p^2, M^2_{\ti t_i}, M^2_{\ti b_j})\big|_{p^2=M_{H^\pm}^2} 
\quad \text{with}\quad 
 \begin{cases} M_{H^\pm}^2\to M_{H^\pm}^2-i M_{H^\pm}\Ga_{H^\pm},\\
M^2_{\ti t_i}\to M^2_{\ti t_i}-iM_{\ti t_i}\Ga_{\ti t_i},\\ M^2_{\ti b_j}\to
 M^2_{\ti b_j}-iM_{\ti b_j}\Ga_{\ti b_j}, \end{cases}\quad i,j =1,2.
\eea
 The required decay widths have been  computed in lowest order
 including all significant two-body decays.

Various cross checks of our calculation have been performed. 
Besides numerical and analytical checks of
UV- and IR-finiteness,
our results were checked versus those obtained by a
independent calculation~\cite{santi} for the real MSSM, 
and very good agreements has been found.

\subsection{Input parameters}
Our calculation is completely general, including all complex phases. 
However, there exist strong constraints on the 
\CP violating parameter space. 
We chose $\mu$  to be zero as default value 
in order to be consistent with the experimental data of the electric 
dipole moments.  The phases of trilinear couplings of the first 
and second generations have marginal effects on the \CP rate asymmetry
because the masses of the corresponding fermions are small. 
In the following, those phases are also taken to be zero. 
The phase of $M_3$, which enters from two loop order, is set to be zero.
The Standard Model input parameters are taken from  \cite{Amsler:2008zzb}.
%
The top mass $m_t= 173.1$ GeV 
is taken from the most recent measurements~\cite{:2009ec}. The
contributions of the CKM-phase to the \CP rate asymmetry are negligible,
 thus the CKM matrix is set to be unit.
For the soft SUSY breaking parameters and $\mu$, we use the following set
as default values (unless specified otherwise),
\bea
\beal \mu &= 200\, \GeV, M_2=200\, \GeV,\, M_3 = 0.8\, M_{\text{SUSY}}, \;  |A_\tau|=|A_t| =|A_b|,\\
M_{\ti Q}&=M_{\ti D}=M_{\ti U}=M_{\text{SUSY}}=500 \,\GeV,\, M_{\ti L}=200 \,\GeV,\, M_{\ti E}= 150\, \GeV. 
\label{eq:parameter1}\eeal
\eea
The values of $\mu$ and  $ M_3$ are chosen as in the $m_h^{\text{max}}$ scenario
 to maximize the lightest neutral Higgs mass \cite{Carena:1999xa}. 
$M_1$ and $M_2$ are chosen as connected via the GUT relation
 $|M_1|= 5/3\tan^2\theta_W |M_2|$. 
Because of this relation, we can 
set $\phi_2 = 0$
while $\phi_1$ is kept as a free parameter. 
The relevant Higgs and SUSY particle masses 
are shown in Table~\ref{table_masses} 
(for $M_{H^\pm} = 300$ GeV and $|A_t| = 800$ GeV). 
Also when varying the parameters, we  have always obeyed the mass constraint
$M_{h_1} > 114.5$ GeV for the lightest neutral Higgs particle
(although for the complex MSSM the limits for the neutral Higgs bosons are less severe
than in the real MSSM) and the experimental limits on the SUSY particles.  
In the following analysis, we will vary 
the trilinear couplings $|A_{\tau/t/b}|$ to show their impact on the asymmetry.
 Since we use
 the $\DRb$ scheme for $\tan\beta$ and the Higgs fields, our results  depend  
on the renormalization 
scale $\mu_{\text{R}}$; more details will be given in section~\ref{sec:scaledependence}.
 We chose $\mu_{\text{R}} = m_{t}$, which is the default value in 
\texttt{FeynHiggs}.

\begin{table}[]
 \begin{scriptsize}
 \bc 
 \caption{\label{table_masses}
  Masses of Higgs bosons and SUSY particles (in GeV) 
    for the parameter set~(\ref{eq:parameter1}) and
    $\phi_1=\phi_\tau =\phi_t=\phi_b=\pi/2$, $|A_t|=800$ GeV.}
\vspace*{0.5cm}
\begin{tabular}{l c c c c c c c c c c c c c c l}
 \hline
$\tan\beta$
& $M_{H^\pm}$ 
& $M_{h_1}$& $M_{\ti \nu}$&$M_{\ti \tau_1}$&$M_{\ti \tau_2}$
& $M_{\ti \chi_1^\pm} $&$M_{\ti \chi_2^\pm} $& $M_{\ti \chi_1^0} $
& $M_{\ti \chi_2^0} $& $M_{\ti \chi_3^0} $& $M_{\ti \chi_4^0} $
& $M_{\ti t_1^0} $& $M_{\ti t_2^0} $& $M_{\ti b_1^0} $
& $M_{\ti b_2^0} $\\
\hline \hline
5& 300& 114.7& 190&155&206&138&272&88&142&208&272&373&645&406&508\\
15&300&120&189&151&209&146&267&89&148&212&226&373&645&448&515\\
\hline 
\end{tabular}\ec
 \end{scriptsize}
\end{table}  
 
\subsection{Dependence on  $\phi_\tau$ and $\phi_1$ }
\setlength{\unitlength}{0.030mm}

\begin{figure}[]
 \begin{center}
\includegraphics[width=7.5cm]{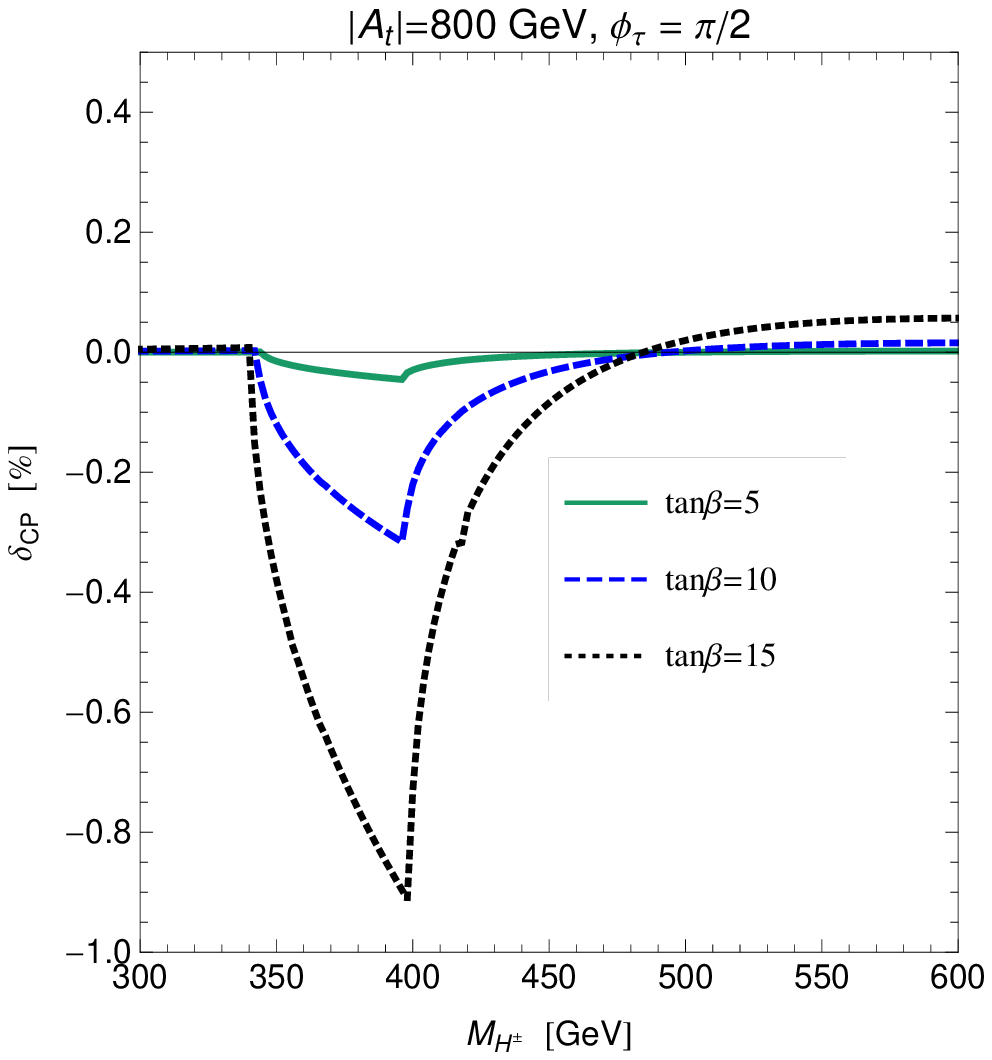}
\qquad
\includegraphics[width=7.5cm]{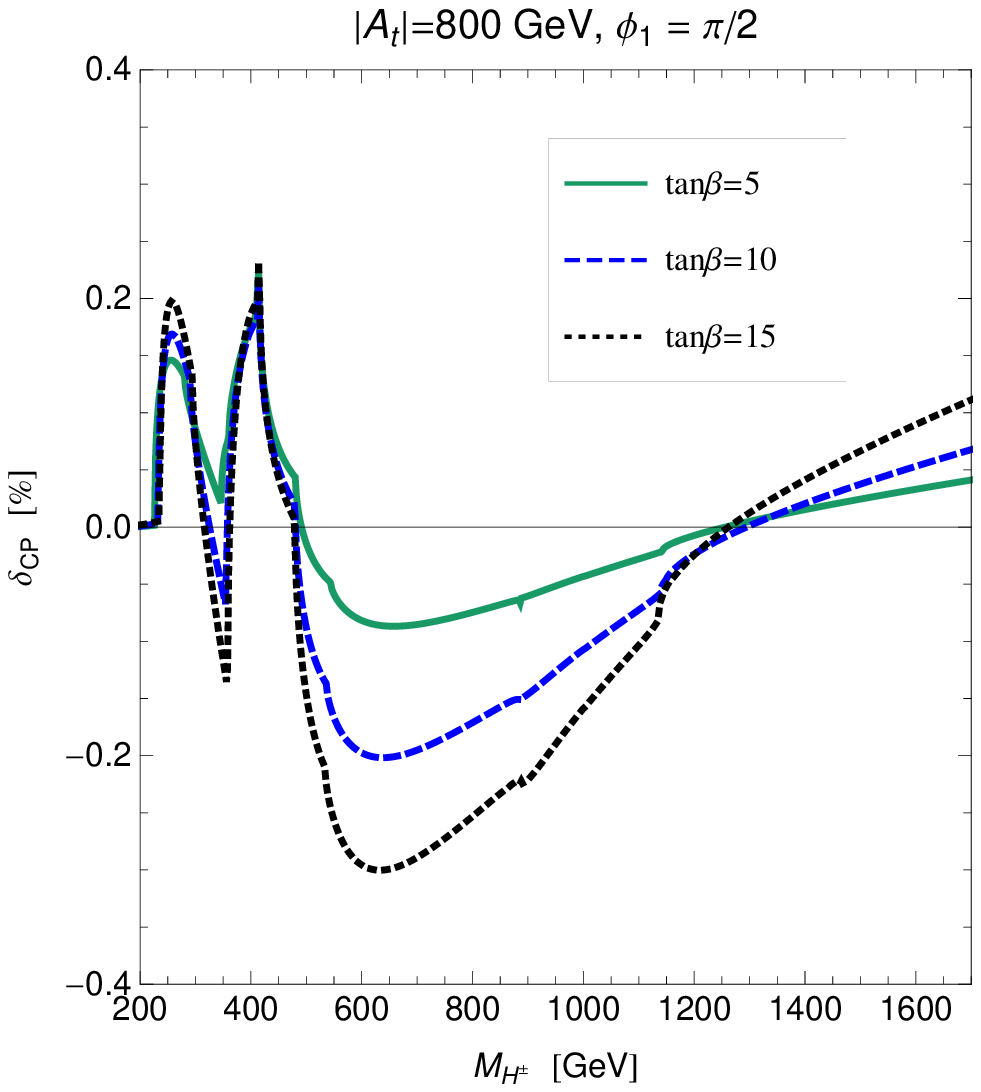}
\caption{$\de_{\CP}$ as function of the charged
 Higgs mass. The left panel is for  $\phi_{\tau} = \pi/2$ 
while the right panel
is for $\phi_1 = \pi/2$. The solid, dashed and dotted lines are 
for $\tan\bet= 5$, 10 and 15, respectively.}\label{fig:numerical0}
\end{center}
\end{figure}
We want to display
the impact of individual phases
on the \CP asymmetry. We therefore keep the phase considered 
 non-zero while 
 all the others are put to zero. The dependence on the phases 
$\phi_{\tau}$ and $\phi_1$ was 
studied already in~\cite{Christova:2003hg}\footnote{
 For a comparison, we have used the same approximation and the same set of
input parameters as in Ref \cite{Christova:2003hg} . Our results
are in agreement with theirs for the case of $\phi_\tau =-\pi/2, \,\phi_1=0$ .
 For the case $\phi_\tau=0,\,\phi_1=-\pi/2$, 
we found a difference resulting from
the coupling between neutral Higgs bosons  and neutralinos,
 $A_{lk}$ in eq.~(A.3) of Ref.~\cite{Christova:2003hg}
where an extra factor $1/2$ is present. 
Adapting this factor,we get  agreement}. 
As mentioned before, we improved 
the calculation by taking  important loop contributions into the
denominator, hence our numerical
results are of two to three times smaller.

For $\phi_\tau =\pi/2$,
  $\de_{\CP}$  as functions of $M_{H^\pm}$ with different values of  
$\tan\bet$ are shown in the left panel of figure \ref{fig:numerical0}.
The diagrams (b, c, f, g) in figure \ref{fig:asymmetry} with $\tilde{\tau}$
and $\tilde{\nu}_\tau$ loops
yield a contribution 
to the CP violating term. Below the 
$\ti\nu_{\tau}\ti\tau_1$ threshold at $M_{H^\pm} \simeq  345\, \GeV$, 
 $\de_{\CP}$ is negligible, 
in spite of contributions from beyond-one-loop terms 
 with the $\bfZ$ factors. The high peaks  correspond to 
the $\ti\nu_{\tau}\ti\tau_2$ 
threshold at $M_{H^\pm} \simeq 396\, \GeV$. Increasing $\tan\bet$ leads to
a rapid decrease of the denominator, owing to
the decreasing tree-level coupling,
which is the main reason for the strongly rising $\de_{\CP}$. 
With $\tan\bet = 5$, the  largest value of 
$\de_{\CP}$  is  about 0.05\%, however with 
 $\tan\bet=15$, $\de_{\CP}$ can go up to 0.91\%. 

For $\phi_1 = \pi/2$, $\de_{\CP}$ is shown in the right panel of 
figure~\ref{fig:numerical0}. 
The diagrams (a, c, d, e) in figure~\ref{fig:asymmetry}, with 
neutralino and chargino loops,  contribute  
to the CP violating term. There are five visible thresholds, 
$\ti\chi^\pm_1\ti\chi^0_1$ at $M_{H^\pm}\simeq 226\,\GeV$, 
$\ti\chi^\pm_1\ti\chi^0_2$ at $M_{H^\pm}\simeq 280\,\GeV$,
$\ti\chi^\pm_1\ti\chi^0_3$ at $M_{H^\pm}\simeq 346\,\GeV$,
$\ti\chi^\pm_1\ti\chi^0_4$ at $M_{H^\pm}\simeq 400\,\GeV$ and
$\ti\chi^\pm_2\ti\chi^0_3$ at $M_{H^\pm}\simeq 480\,\GeV$. 
$\de_{\CP}$ can reach 0.3\% above the $\ti\chi^\pm_1\ti\chi^0_1$ threshold,
in general, however, it is rather small.

\subsection{Dependence on $\phi_t$ and $\phi_b$}
\begin{figure}[]
 \begin{center}
\subfloat[ ]{\includegraphics[width=8cm]
{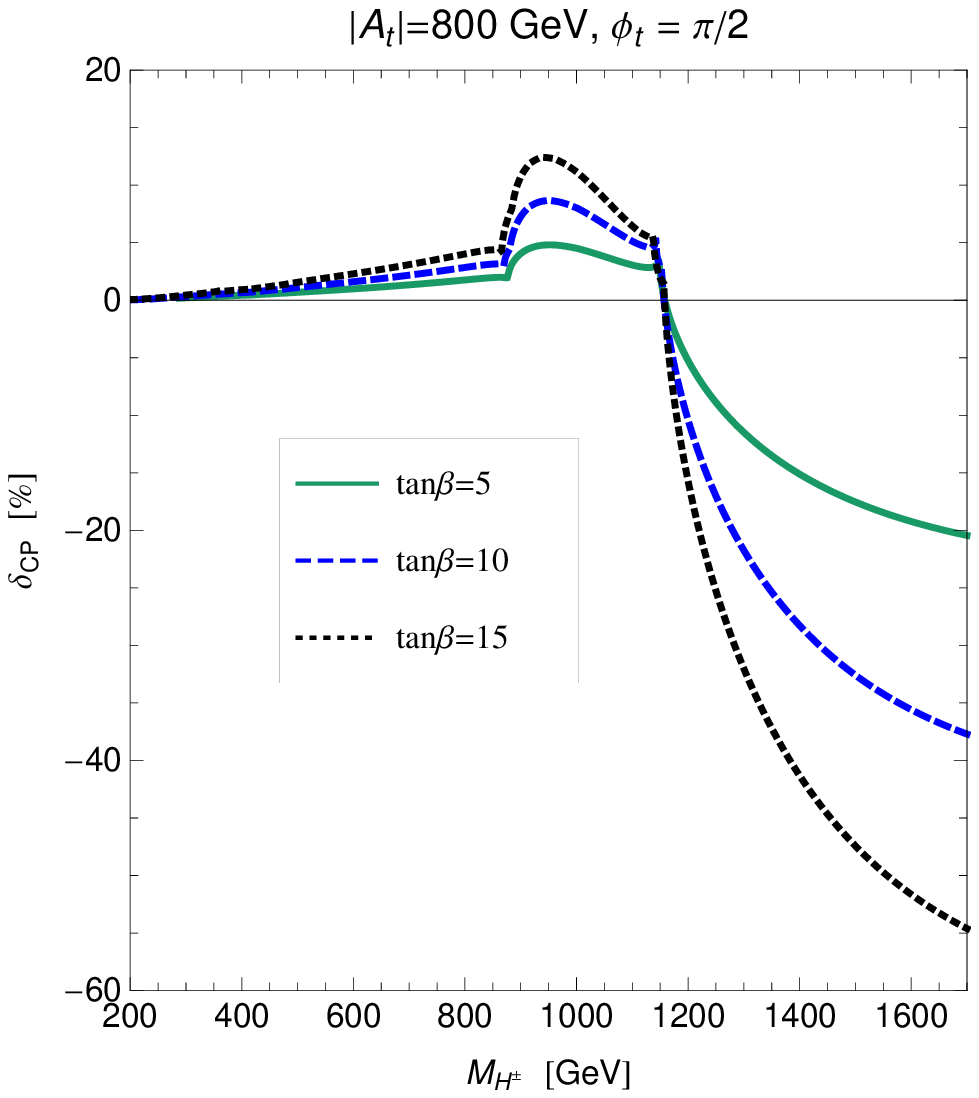}\label{fig:top_2_A800}}\quad
\subfloat[]
{\includegraphics[width=8cm]{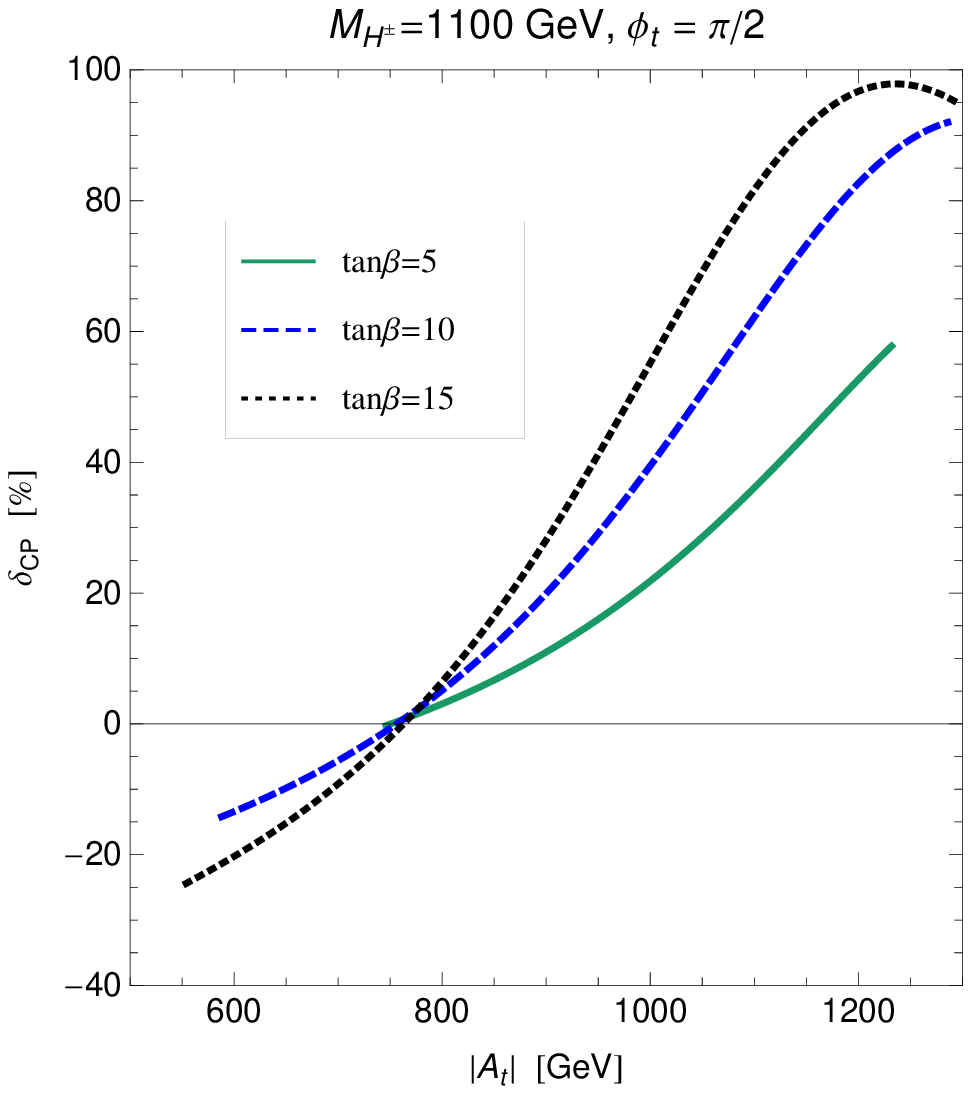}\label{fig:top_2_MHp1100}}
\caption{The \CP asymmetry  as functions (a) of the charged
 Higgs mass,  (b)  of $|A_t|$.
The solid,  dashed and dotted lines are for $\tan\bet=5,\,10$ and 15,
respectively.}\label{fig:numerical1}
\end{center}
\end{figure}

Significantly larger values of $\de_{\CP}$ can occur when $\phi_t$ and $\phi_b$ are 
non-zero and the CP violating terms get contributions from diagrams with
top and bottom squarks loops (figure~\ref{fig:asymmetry}). 
The left panel of figure 
\ref{fig:numerical1} shows the  CP asymmetry as a function of the charged Higgs
 mass for $\phi_t = \pi/2$. There are two visible thresholds, 
 $\ti t_1\ti b_1$ at $M_{H^-}\simeq 873\, \GeV$ 
and $\ti t_2 \ti b_2$ at $M_{H^-}\simeq 1149\,\GeV$ for 
$\tan\beta = 5$. 
\setlength{\unitlength}{0.030mm}

The CP asymmetry is sizeable 
both for $M_{H^\pm}$  below and above 
the $\ti t_1\ti b_1$ threshold, especially for larger values of
$\tan\beta$.
Below the $\ti t_1\ti b_1$ threshold, the most important 
term contributing to the CP asymmetry is the interference 
between diagram (c) in figure \ref{fig:asymmetry}
and the triangles with top and bottom quarks.  
Close to  the threshold, the interference 
of the diagrams (b, f, g) in figure \ref{fig:asymmetry} and the 
tree diagram  are dominant. 
We observe that the individual contribution
from the  H-W mixing diagrams 
and the triangles  with same particles inside loops can be much  
larger than the Born-term  at the $\ti t_i\ti b_j$ thresholds.
However, they carry opposite signs
and are almost of the same order of magnitude. 
The sum of both can be comparable with the 
Born term and is very sensitive with respect to 
$\phi_t$, $|A_t|$ and $\tan\bet$. 

Above the $\ti t_1\ti b_1$ threshold, $\de_{\CP}$ 
can become very large. 
It can rise up to -51.6\% at $M_{H^-}$=1600 GeV, $\tan\bet$=15.
This is a common feature of 
charged Higgs decays, as mentioned in 
Ref\cite{Christova:2002ke}.
Moreover, $\de_{\CP}$ has a strong dependence on $|A_t|$, 
as one can see in 
the right panel of figure~\ref{fig:numerical1}. 
The $|A_t|$ range is compatible with $M_{h_1} > 114.5\, \GeV$. 

\begin{figure}[]
 \begin{center}
\includegraphics[width=9cm]{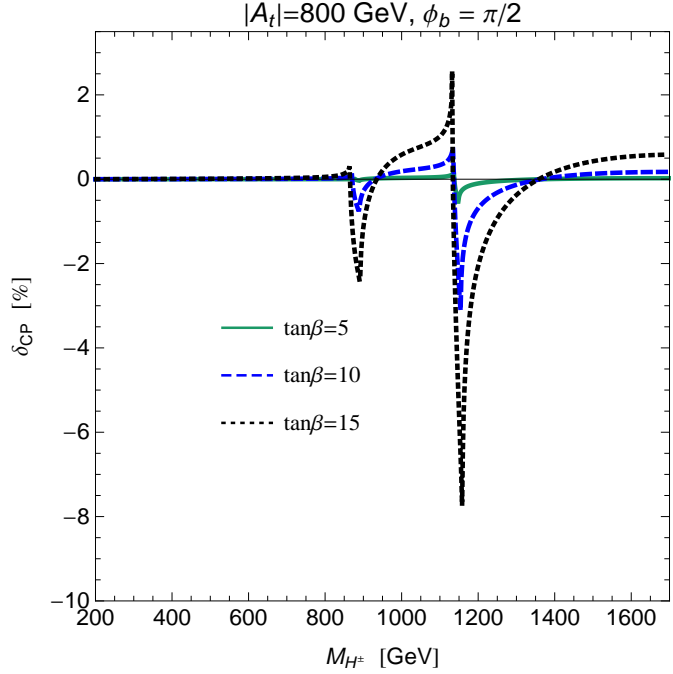}
\caption{The \CP asymmetry as function of 
charged Higgs mass, for $\phi_b=\pi/2$. The solid, dashed and 
dotted lines are for $\tan\bet=5,\,10$ and 15,  respectively.}
\label{fig:numerical2}
\end{center}
\end{figure}

The impact of the phase $\phi_b$ on $\de_{\CP}$ is shown in 
figure~\ref{fig:numerical2}.
It can be sizeable above $M_{H^-}$ around the $\ti t_1\ti b_1$ threshold,
however it is still small compared to the effect of the phase $\phi_t$.
For $|A_t|=800\,\GeV$,
the largest value of $\de_{\CP}$ obtained for
 $\tan\bet=15 $ is about 8\% close to the $\ti t_2\ti b_2$ threshold.
\setlength{\unitlength}{0.030mm}

\begin{figure}[]
 \begin{center}
\subfloat[ ]{\includegraphics[width=7.5cm]
{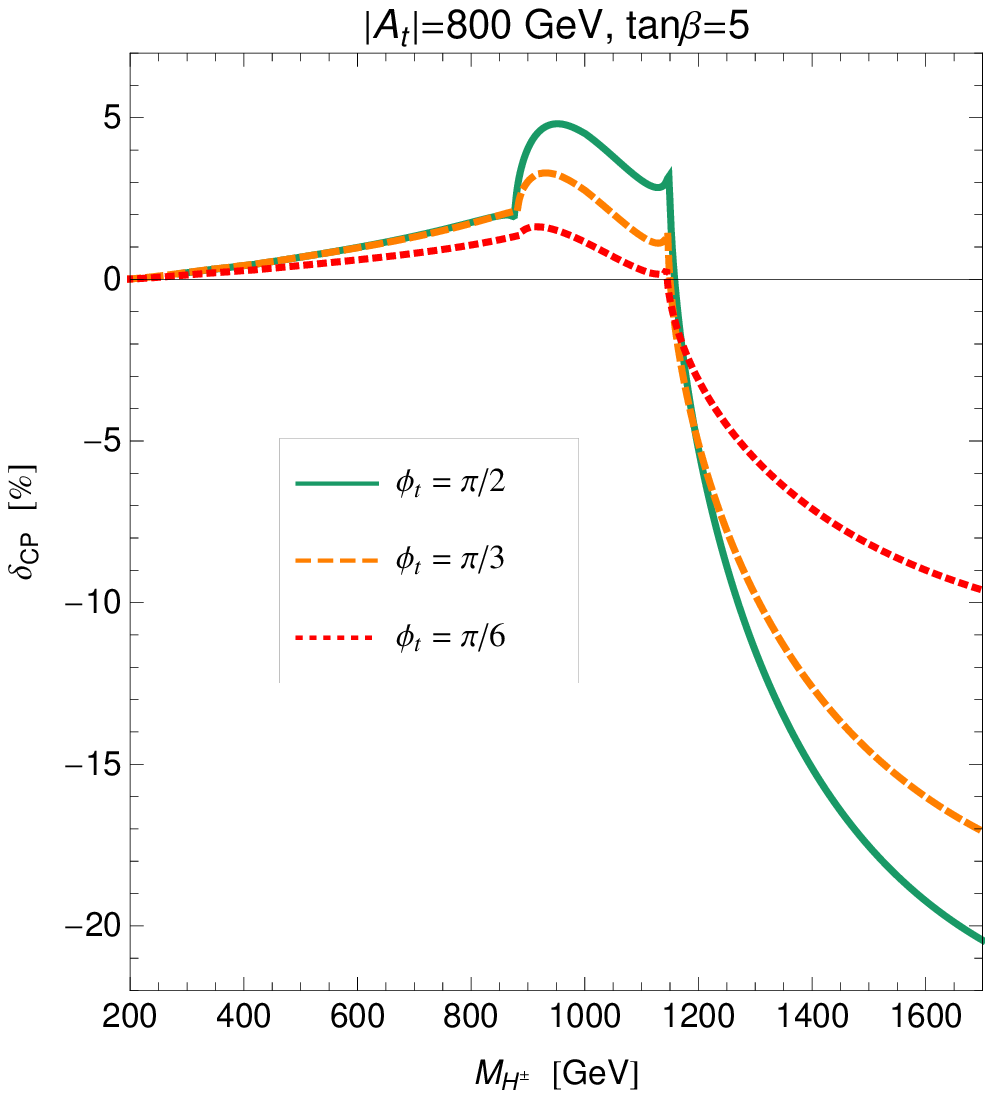}\label{fig:phistop_2_3_6}}\quad
\subfloat[ ]
{\includegraphics[width=7.5cm]{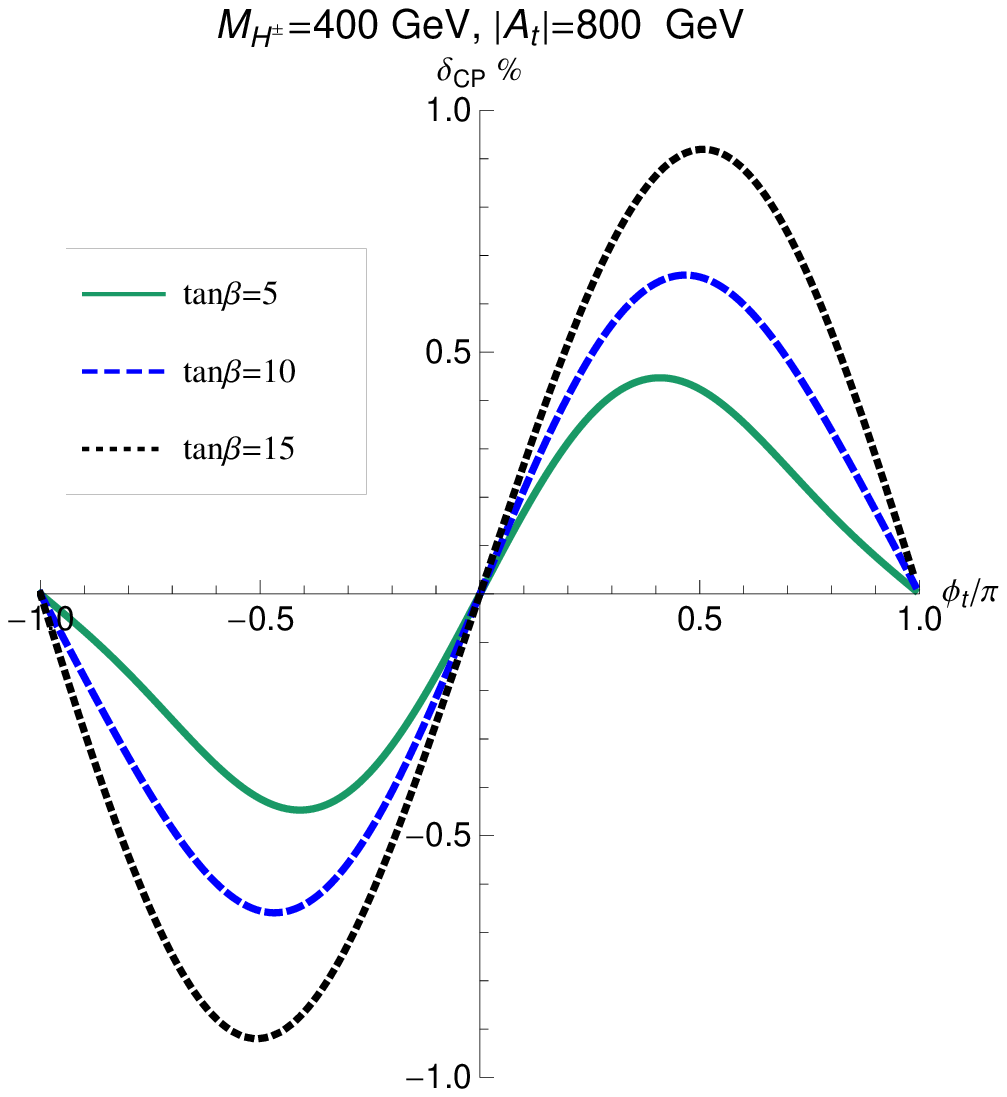}\label{fig:tb5_10_15_stop}}
\caption{The  CP asymmetry as function (a) of the
charged Higgs mass for different values of 
$\phi_t = \{\fr\pi 2,\, \fr\pi 3,\, \fr \pi 6\}$ 
(b) of the  CP asymmetry as functions
 the phase $\phi_t$ for $\tan\bet= 5,\, 10,\, 15$.}
\label{fig:numerical3}
\end{center}
\end{figure}

The  dependence of the \CP asymmetry on the phase 
of $A_t$ is illustrated 
in figure \ref{fig:phistop_2_3_6}, 
where we present $\de_{\CP}$ as a function of the charged Higgs mass with
different values of $\phi_t = \fr\pi 2,\, \fr\pi 3,\, \fr \pi 6$.
Figure~\ref{fig:tb5_10_15_stop} shows the CP  asymmetry at
 $M_{H^-}= 400$ GeV  as a function of phase $\phi_t$ with
  $\tan\bet = 5,\, 10,\, 15$. For $\tan\bet=15$ the maximum is at 0.92\%
for $\phi_t = 0.51\pi$. Compared to the contributions from $\phi_1$ and 
$\phi_{\tau}$ at low values of $M_{H^-}$, the impact of $\phi_t$ 
on $\de_{CP}$ is considerably  bigger, although not very strong from the
absolute numbers.
\begin{figure}[]
 \begin{center}
\subfloat[ ]{\includegraphics[width=7.5cm]
{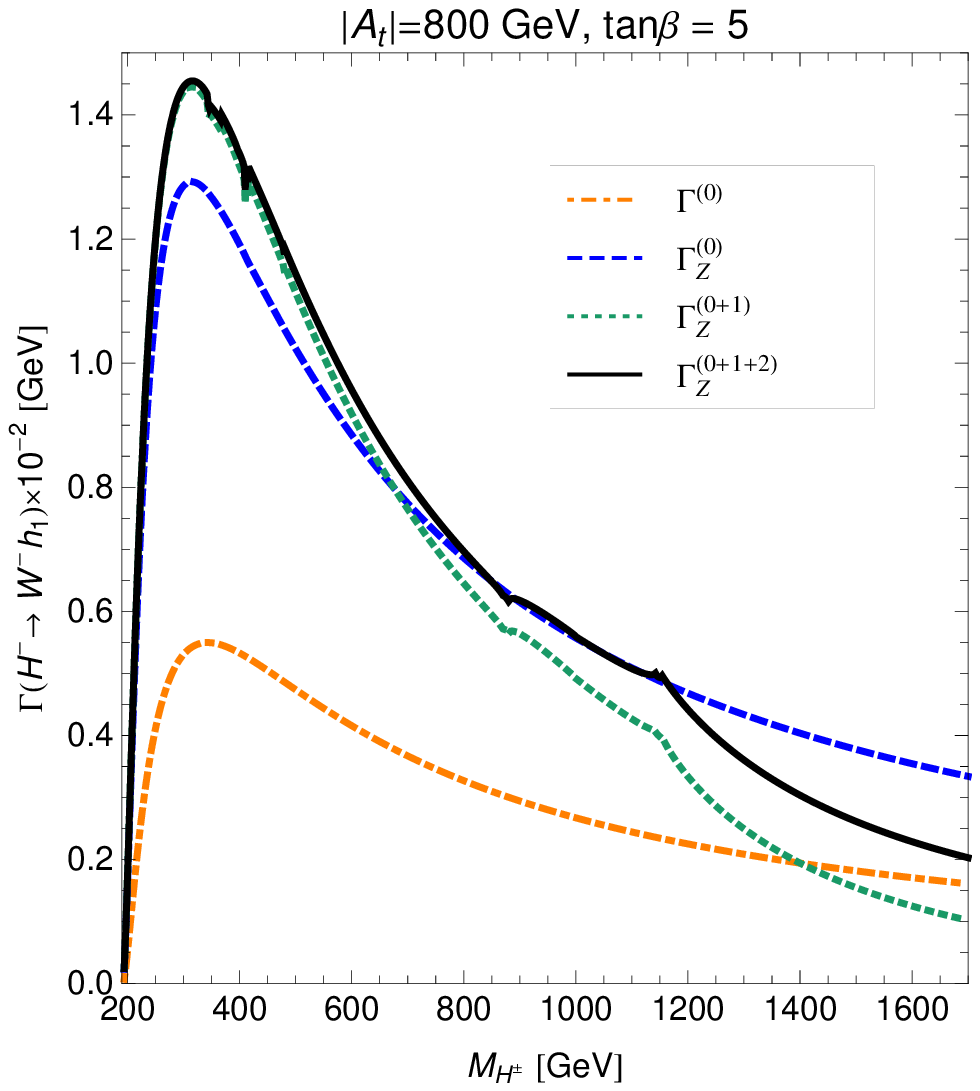}\label{fig:width_TB_5_800}}\quad
\subfloat[ ]
{\includegraphics[width=7.5cm]{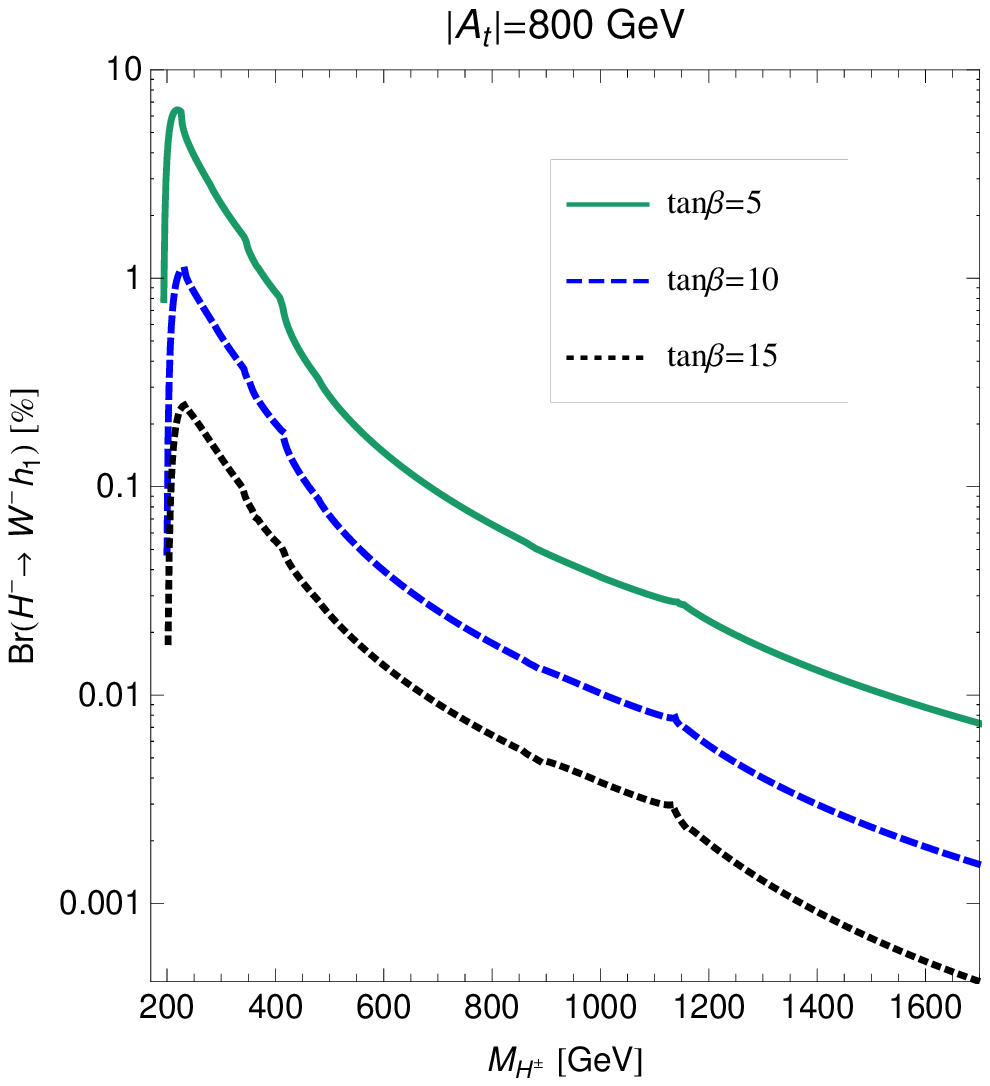}
\label{fig:BRratio}}
\caption{(a) The  Born, improved Born, improved one-loop and 
full decay widths
corresponding to dot-dashed, dashed, dotted and solid lines  as functions
 of the charged Higgs boson mass. 
(b) The branching ratios of the decay $H^-\to  W^-h_1$ as functions
of $M_{H^-}$, 
for $\phi_1=\phi_{\tau}=\phi_t =\phi_b=\pi/2$.}
\label{fig:numerical4}
\end{center}
\end{figure}

As already mentioned, the sum of the decay widths 
for $H^\pm\to W^\pm h_1$ is an important ingredient for $\de_{\CP}$
and the Born approximation is in general insufficient.
Therefore we address here the 
decay widths and branching ratios 
and the higher-order effects. For illustration we choose the 
decay $H^-\to W^-h_1$.
In figure~\ref{fig:width_TB_5_800}, we show the Born, improved Born,
improved one-loop and full decay widths, 
as described in section~3.
The improved Born and improved one-loop decay widths are
defined in~(\ref{eq:improveborn}) and (\ref{eq:improveloop}). We choose 
$\phi_1=\phi_{\tau}=\phi_t =\phi_b=\pi/2$ for this analysis. 
For $M_{H^\pm} = 300\,\GeV$, the one-loop vertex corrections can go up
 to 12.4\% 
while at $M_{H^\pm} = 1.6\, \TeV$  corrections reduce to -35.4\% compared to 
improved Born result. 
For low $M_{H^\pm}$, the improved one-loop and 
the full result are quite close to each other, 
but around and above the $\ti t_1\ti b_1$ threshold,
the full result is  clearly larger.

 In  figure \ref{fig:BRratio}, we show the branching ratio of the decay 
$H^-\to h_1 W^-$  for different values of $\tan\bet$,
using the full decay width. The other 
relevant decays of the charged Higgs boson are computed in lowest order. For
$\tan\bet =5$,  the branching ratio  can reach  6.4\% 
at $M_{H^\pm}\simeq  219\, \GeV$. 
Around this point, 
the charged Higgs can decay  mainly to $t\,b$ and $\tau\,\nu_\tau$. When the
mass of  charged Higgs mass increases, the channels to charginos and neutralinos,
stop and sbottom open. Thus, the branching ratio of $H^-\to h_1 W^-$ drops 
rapidly, which makes 
it difficult to access  $\de_{\CP}$ experimentally.
The branching ratio also depends strongly on the value of $\tan\bet$,
especially for low values of $\tan\bet$,
where the channels  $H^\pm\to h_1 W^\pm$ are interesting. 

\subsection{Dependence on $\phi_\mu$}
\begin{figure}[]
 \begin{center}
\includegraphics[width=9cm]{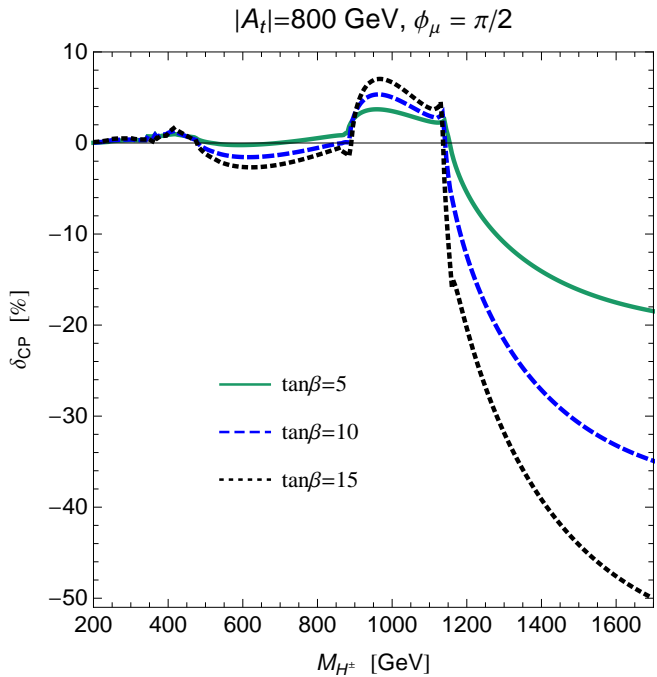}
\caption{The \CP asymmetry as functions of 
charged Higgs mass, for $\phi_\mu=\pi/2$. The solid, dashed and 
dotted lines are for $\tan\bet=5,\,10$ and 15,  respectively.}
\label{fig:phimudependence}
\end{center}
\end{figure}

The phase  of  $\mu$ is severely constrained by the experimental limits on the electric dipole moments
of electron and neutron. This bounds can, however, be circumvented by a specific
fine-tuning of the phases of $\mu$ and of the non-universal SUSY 
parameters~\cite{Ibrahim:1997gj}, leaving room also for a large phase $\phi_\mu$.
We thus illustrate the effect of a large $\phi_\mu$
on $\de_{\CP}$ in 
Figure~\ref{fig:phimudependence}, which displays $\de_{\CP}$ as a function
of $M_{H^\pm}$ for $\phi_\mu=\pi/2$. 
The \CP violating part receives contributions from all diagrams in
figure~\ref{fig:asymmetry}. 
For charged Higgs boson masses below the $\ti t_1\ti b_1$ threshold, the main
contribution to $\de_{\CP}$ comes from the neutralino-chargino loops; above the threshold it is 
again dominated by the $\ti t_1\ti b_1$ loops.

\subsection{Scale dependence} \label{sec:scaledependence}
\begin{figure}[]
 \begin{center}
\includegraphics[width=9cm]{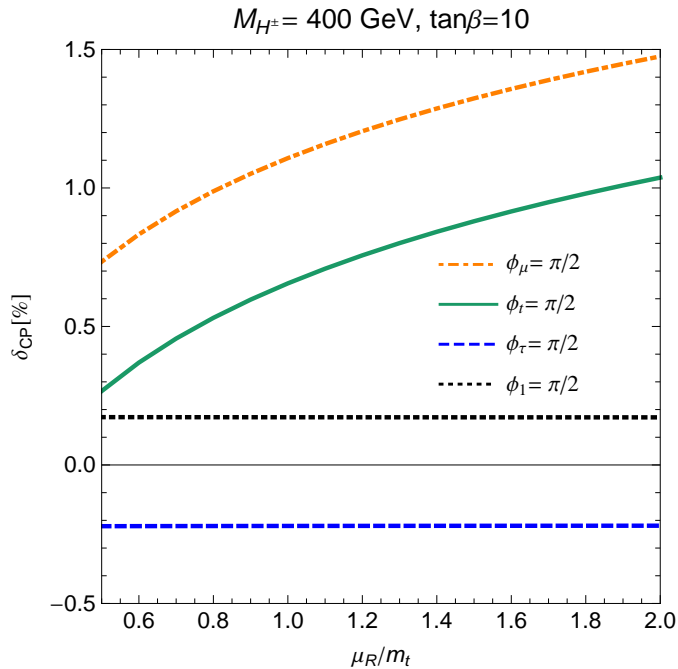}
\caption{The CP asymmetry as functions of the renormalization scale for different phases. 
$\mu_R$ is varied in the range $[m_t/2,2m_t]$.}
\label{fig:scaledependence}
\end{center}
\end{figure}

Here we comment on the dependence of the CP asymmetries on the
renormalization scale $\mu_R$. Choosing a concrete example,  
Figure~\ref{fig:scaledependence}  shows $\de_{\CP}$ versus
of $\mu_R$ at $M_{H^\pm} = 400$ GeV and $\tan\beta =10$.  
The dependence of $\de_{\CP}$
on $\mu_R$ comes mainly from the CP violating contribution in the numerator of~(\ref{eq:asymmetry}).     
The strict one-loop contribution to the CP violating part of the decay width  
does not depend on $\mu_R$
since it arises from the imaginary part of one-loop integrals. 
We however consider also higher-order terms, like
the Higgs-mixing term $\bfZ_{hA}M_A^{\tree}\de M_h$, which depends on $\mu_R$
through the $\bfZ$ factors from the Higgs renormalization.
For $\phi_\tau$ and $\phi_1$, such terms are negligible  and
the dependence on $\mu_R$ is irrelevant.
For $\phi_\mu$ and $\phi_t$ they are more important,
as one can see in the figure.
For $M_{H^\pm}$ values above the  $\ti t_1\ti b_1$ 
threshold, the one-loop contribution is the most important, and then the 
$\mu_R$ dependence is much weaker.  
\subsection{The CPX scenario}
\begin{figure}[!ht]
 \begin{center}
\subfloat[ ]{\includegraphics[width=7.5cm]
{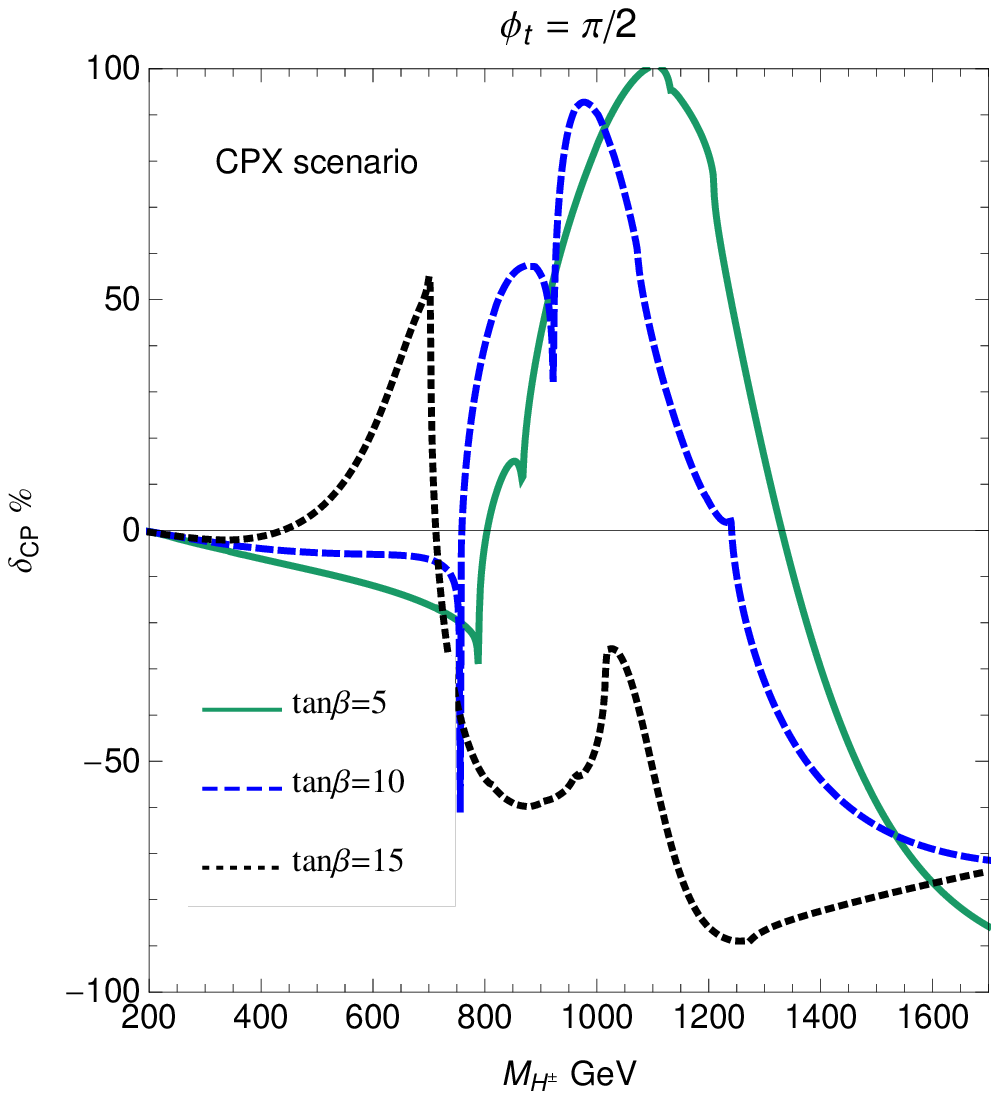}\label{fig:top_2_CPX}}\quad
\subfloat[ ]
{\includegraphics[width=7.5cm]{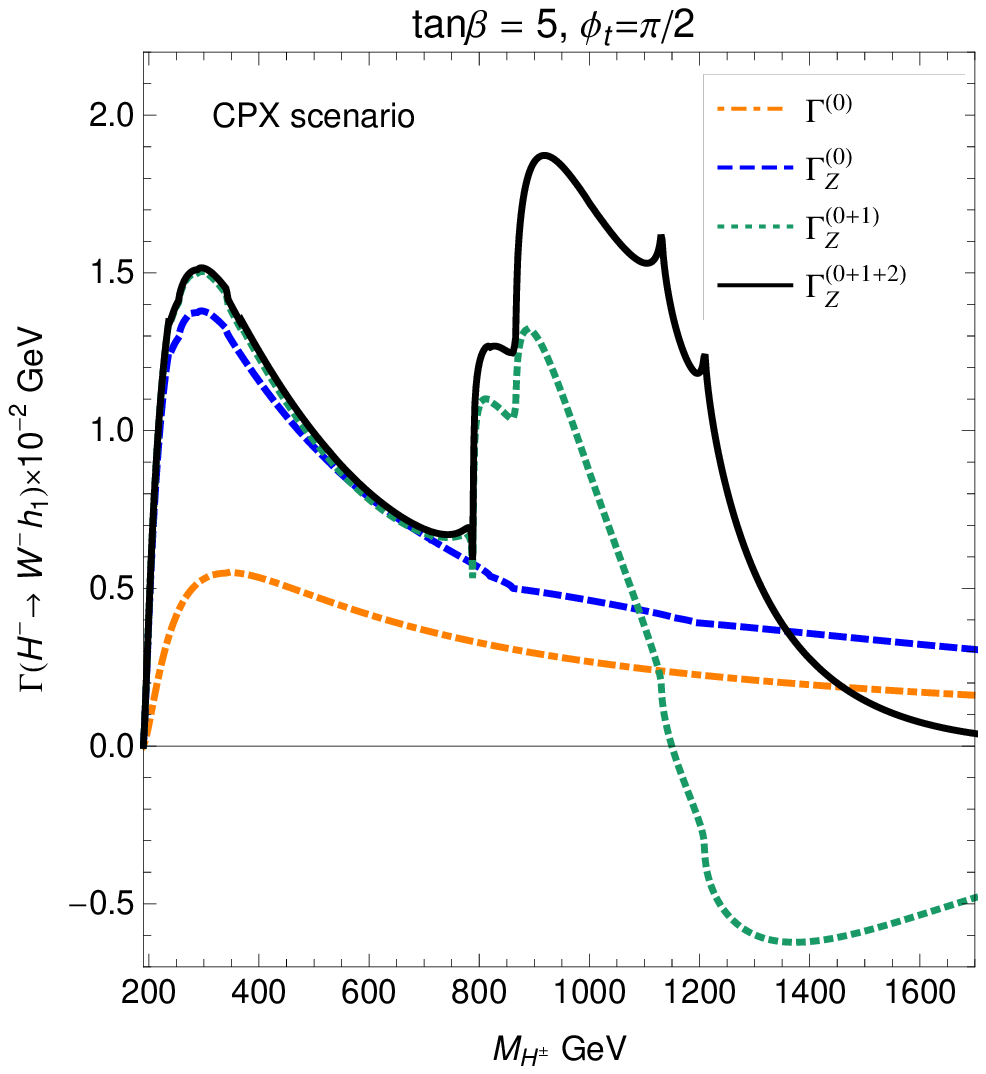}
\label{fig:width_CPX}}
\caption{(a) $\de_{\CP}$ as function of the charged Higgs mass in the CPX scenario.
 (b) The  Born, improved Born, improved one-loop and 
full decay widths,
corresponding to dot-dashed, dashed, dotted and solid lines,  as functions
 of the charged Higgs mass in the CPX scenario.}
\label{fig:numerical5}
\end{center}
\end{figure}

A case of particular interest is the CPX scenario where the 
SUSY parameters
maximize the CP-violating effects
due to the large value of the product $\Img(\mu A_t)/M_{\text{SUSY}}^2$ 
\cite{Carena:2000ks}. According to
Ref.~\cite{Williams:2007dc}, we use the following 
set of on-shell parameters
\bea \beal \mu &= 2000\,\GeV, M_{\text{SUSY}} = 500 \, \GeV, |A_f| = 900\,\GeV, 
\\
M_3 &=1000\,\GeV, M_2= 200\,\GeV,  M_1= 5/3\tan^2\theta_W M_2. \eeal \eea
In figure~\ref{fig:top_2_CPX}, we display the CP asymmetry caused by the 
complex phase of $A_t$ for $\tan\bet = 5,\, 10,\, 15$. As one can see,
$\de_{\CP}$ is quite large both below and above $\ti t_1\ti b_1$ threshold.
For $\tan\bet = 5$, $\de_{\CP}$ is about -6\% at $M_{H^\pm} \simeq 400\,\GeV$ 
and can reach 100\% at $M_{H^\pm} \simeq 1116\,\GeV$. 
In figure~\ref{fig:width_CPX}, 
the decay width is shown as function of $M_H^\pm$. 
Note that above the  $\ti t_1\ti b_1$ threshold, the one-loop correction
becomes very large, making the improved one-loop width negative,
which demonstrates that this kind of approximation is unphysical and shows the
importance of not truncating the squared amplitude.

\section{Conclusions}
We have  calculated the \CP violating asymmetry from the
decays $H^\pm\to h_1W^\pm$ originating from non-vanishing complex phases
in the complex MSSM. 
All the phases that can give sizable contributions to $\de_{\CP}$ are
taken into account and discussed.  
The impact of the phases $\phi_{\tau}$, $\phi_1$ and 
$\phi_b$ on \CP rate asymmetry is of some significance only above
the threshold.  The phase $\phi_t$
can yield large contributions to 
the  \CP asymmetry both below and above the thresholds.  $\phi_t$ and  $\phi_\mu$
can induce large $\de_{\CP}$ at large $M_{H^\pm}$.
$\de_{\CP}$ depends strongly on $M_{H^\pm}$, $|A_{t}|$
and $\tan\bet$.

We have also presented the decay width  and
the branching ratio of the decay  $H^-\to h_1W^-$.
They turn out to be significant in particular for small values 
of $\tan\bet$ and low masses of the charged Higgs boson. With increasing
mass they become rather small.


Although the CP asymmetry can be large, the small branching ratios make
the experimental observability quite difficult.
A characteristic number for a feasibility estimate is the quantity
$N= (\de_{\CP}^2 \times \text{Br})^{-1}$ \cite{Eilam:1991yv}, 
the number of the (at least) required charged Higgs bosons
to be produced for observing the CP asymmetry.
For $M_{H^\pm}=500$~GeV and an asymmetry of $-9\%$, as in the CPX scenario for $\tan\beta=5$
with a branching ratio of 4.2\%, one would need about $N=3\cdot 10^3$.
At the  Large Hadron Collider (LHC) the dominant production occurs through the
partonic process $gb\to tH^-$ (see e.g.~\cite{Djouadi:2005gj} for a review),
which with a cross section of 19~fb could provide 
such a number of charged Higgs bosons for an integrated luminosity
of 160~fb$^{-1}$.  
Considering a very large asymmetry of $0.9$ as for $M_{H^\pm}=1000$ GeV,    
one has to cope with a very small branching ratio of $6.7\cdot 10^{-4}$, requiring
$N= 1.9\cdot 10^3$; for a production cross section of 1.2~fb a luminosity of more than
$1.6\cdot 10^3$ fb$^{-1}$ would be needed, which is outside the scope of the 
LHC with the envisaged design luminosity (but might be of interest for an upgraded SLHC).

For a more realistic study, moreover, one has to take into account that CP violating effects 
are also part of the main production processes  
$gb(\bar{b})\to t H^- \, (\bar{t} H^+)$~\cite{Christova:2008jv}, which
makes a complete calculation for $H^\pm$ production and decay at NLO necessary.   
  
At a Linear Collider, the basic production process $e^+e^-\to H^+H^-$ has the advantage
of providing a symmetric state, from which the observation of CP violation in the 
charged Higgs decays might look more promising, but is also depleted by low production rates
and branching ratios.
The cross section for pair production with $M_{H^\pm} = 500$ GeV
at a center-of-mass energy of 3 TeV (CLIC) is  2.6~fb,  
which would require an integrated luminosity of $1.2\cdot 10^3$~fb$^{-1}$.
For lower Higgs masses (up to 400 GeV), one can expect higher production rates
at a 1 TeV collider, but the predicted CP asymmetries are rather small in that range.

\subsection*{Acknowledgements}
We would like to thank Santiago Bejar Latonda and  David Lopez Val
for cross-checking some parts of our calculations and
Le Duc Ninh  and  K.E. Williams for fruitful discussions.

\end{document}